\documentstyle[prb,epsfig,aps]{revtex} 
\begin{document}
\draft
\title{The order-disorder character of the $(3 \times 3)$ to 
$(\sqrt 3 \times \sqrt3)$R30$^{\circ}$ phase transition 
of Sn on Ge(111)}

\author{
L. Petaccia,$^{1}$  
L. Floreano,$^{1,}$\footnote{Corresponding Author:
Luca Floreano,
Surface Division,
Laboratorio TASC-INFM,
Basovizza, SS14 Km 163.5
I-34012 Trieste,
Italy.
E-mail: floreano@sci.area.trieste.it}
A. Goldoni,$^{2}$
D. Cvetko,$^{1,2,3}$
A. Morgante,$^{1,4}$
L. Grill,$^{1}$
A. Verdini,$^{1}$
G. Comelli,$^{1,4}$
G. Paolucci,$^{2}$ and 
S. Modesti$^{1,4}$
}

\address{
$^{1}$ Laboratorio TASC, Istituto Nazionale per la Fisica della Materia, 
Basovizza SS14 Km 163.5, I-34012 Trieste, Italy}

\address{
$^{2}$ Sincrotrone Trieste SCpA, Basovizza SS14 Km 163.5, I-34012 Trieste, 
Italy}

\address{
$^{3}$ Jo{\v z}ef Stefan Institute, Department of Physics, 
Ljubljana University, Ljubljana, Slovenia}

\address{
$^{4}$ Department of Physics, University of Trieste, Via 
Valerio 2, I-34100 Trieste, Italy}

\narrowtext
\onecolumn

\maketitle
\begin{abstract}

The $\alpha$-phase of Sn/Ge(111) has been investigated from 120~K up to 
500~K, 
using synchrotron radiation core level photoemission. 
By means of photoelectron diffraction experiments, we 
verified that the rippled structure of the 
low temperature $(3 \times 3)$-phase 
is preserved in the 
$(\sqrt 3 \times \sqrt3)$R30$^{\circ}$-phase at room temperature, 
thus confirming the  
order-disorder character of the phase transition.
We also found that at least two 
components are present in the Sn 4d core level spectra up to 500~K, 
i.e about 300~K 
above the onset of the transition from the low temperature 
$(3 \times 3)$-phase to the 
$(\sqrt 3 \times \sqrt3)$R30$^{\circ}$-phase, thus excluding the 
occurence of any displacive transition. 

\end{abstract}
\vspace{1cm}


\narrowtext
\twocolumn

The system formed by one third of a monolayer (ML) of Sn adatoms atop the 
T$_{4}$ sites of the Ge(111) surface ($\alpha$-phase) 
exhibits a transition from a low temperature (LT) $(3 \times 3)$ phase 
to a high temperature (HT) 
$(\sqrt 3 \times \sqrt3)$R30$^{\circ}$ phase. 
This transition is gradual, reversible and has a critical temperature 
T$_{c} \sim$~220~K. 
The phase transition was first attributed to 
the manifestation of a commensurate surface charge density wave state in the LT 
phase stabilized by many body effects, in particular by electron 
correlation.\cite{carpinelli,carpinelli2,santoro,goldoni}
Later on, an order-disorder model,\cite{uhrberg} invoking dynamic 
fluctuations as microscopic driving mechanism,\cite{avila} 
has been proposed to explain the published data. 
The research then focused on the role of surface 
defects\cite{melechko,weitering} and the 
determination of the surface structure\cite{bunk,zhang,floreano,petaccia} 
in an attempt to discriminate 
between the two interpretations.

At present, the $(3 \times 3)$ structure has been determined with 
a good agreement between  surface X-ray diffraction\cite{bunk,zhang} 
(SXRD) and photoelectron diffraction\cite{petaccia} (PED) measurements.
At low temperature, 
the Sn adatoms occupy inequivalent $T_{4}$ sites, with one Sn adatom 
(out of three  within the unit cell) 
protruding above the surface. This inequivalency is also confirmed by 
photoemission experiments, where at least two components 
(with intensity ratio close to 1:2) are required to 
fit the Sn 4d spectrum properly.\cite{uhrberg,floreano,petaccia,lelay,avila99}

For the structure of the 
$(\sqrt 3 \times \sqrt3)$R30$^{\circ}$ phase 
at room temperature (RT), only SXRD measurements are 
available.\cite{bunk,zhang} They
slightly favor a structure with equivalent Sn adatoms, thus
suggesting that the transition is indeed due to a 
pseudo-Jahn-Teller effect,\cite{bunk} i.e. it would be a displacive 
phase transition. This result is hardly conciliating with a very 
recent He diffraction study of the $(3 \times 3)$ order parameter, 
which indicates the occurence of a 3-state Potts order-disorder phase 
transition at $T_{c} \sim 220$~K.\cite{floreano2} 
Moreover, the hypothesis of a displacive transition at 220~K 
is also not compatible with the core level  photoemission 
experiments, where the inequivalency between the Sn atoms 
seems to remain also in the 
$(\sqrt 3 \times \sqrt3)$R30$^{\circ}$ phase, since the Sn 4d 
lineshape does not change significantly from $\sim$~100-150~K up to room 
temperature.\cite{uhrberg,avila,lelay,uhrberg2} 
This discrepancy urges for PED structural measurements of the RT phase.

The hypothesis of a displacive transition, being 
recently suggested to be driven by a phonon softening,\cite{perez}   
asks for 
experiments in a more extended temperature range.
To our knowledge no data
have been published at temperatures higher than 300~K, while it is well known 
that the displacive phase transitions are always accompanied by an intermediate 
disordered stage around the critical 
temperature,\cite{bruce,tosatti,toennies} manifesting the 
change in surface corrugation only at a temperature substantially
 higher than T$_{c}$. 
It is therefore of 
utmost importance to have a clear indication of the order-disorder character of 
the transition around T$_{c}$ and to know if the two components observed in the Sn 
4d core level spectra at LT and RT are also present far from T$_{c}$,
 up to the 
highest temperature where this phase is stable 
($\sim$~550~K).\cite{ichikawa}

By means of photoelectron diffraction (PED) measurements, we have 
already shown that in the $(3 \times 3)$ phase the observed splitting 
of the Sn 4d core levels in two 
components derives from adatoms with different bonding 
configuration.\cite{floreano}
A vertical ripple of  0.3~\AA~ between the Sn adatoms has been estimated 
by simulations, with one 
adatom out of three that protrudes above the 
surface. In particular, the adatom at the higher 
height level is associated to the Sn 4d component with the larger 
binding energy.\cite{petaccia}
 The three nearest-neighbor Ge atoms follow the Sn adatom vertical 
distortion, i.e. their bond angle and length remain almost unchanged. 
This rippled structure is 
in good agreement with both the SXRD experiments\cite{bunk} and recent density-
functional theory (DFT) calculations\cite{degironcoli,ortega} 
for the LT $(3 \times 3)$-phase.
 
In this paper, we report on the evolution of the Sn 4d core levels and on their 
PED patterns at different temperatures. 
The measurements were performed at the ELETTRA Synchrotron facility in 
Trieste, in the ultra high vacuum end-stations of the 
ALOISA\cite{alo99,aloweb} and SuperESCA\cite{superesca} beamlines. 
The Sn 4d core levels photoemission spectra presented here were 
measured at the SuperESCA beamline with an overall energy resolution better 
than 120 meV. The PED experiments were performed at the ALOISA beamline. 

Details about the sample preparation and PED experiment at  ALOISA 
 are given elsewhere.\cite{petaccia} 
At the SuperESCA beamline, the 
$\alpha$-phase was prepared by dosing 1/3~ML of Sn on a clean 
Ge(111)-c$(2 \times 8)$ surface 
kept at RT. The optimum coverage and annealing treatment 
for the $(\sqrt 3 \times \sqrt3)$R30$^{\circ}$ 
and $(3 \times 3)$ phases was checked by optimizing the intensity 
of the LEED $(3 \times 3)$ 
extra spots of the LT phase and the intensity of the Sn induced surface band at 
the Fermi level in the photoemission spectra. This feature is indeed very sensitive 
to the quality of the $(\sqrt 3 \times 
\sqrt3)$R30$^{\circ}$ phase.\cite{goldoni,uhrberg2} 
We have investigated also the 
$(2 \times 2)$ phase of Sn/Ge(111), obtained by dosing $\sim$~0.2~ML of Sn and annealing at 
450~K. The temperature stability during measurements was better than $\pm$~10~K.

\begin{figure}[tbp]
	\centering
	\includegraphics{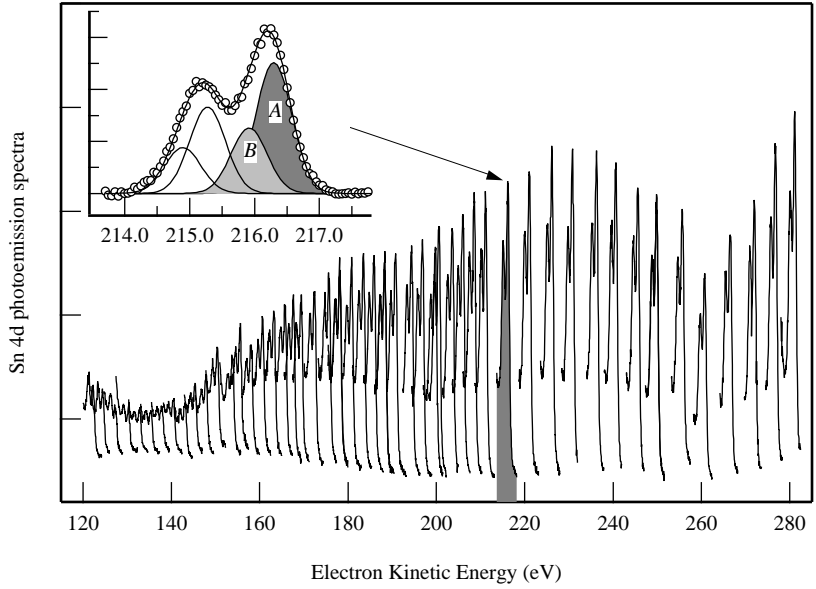}
	\caption{Sn 4d photoemission spectra taken in normal emission from 
	the $(\sqrt 3 \times \sqrt3)$R30$^{\circ}$ phase at RT. Each 
	spectrum has been taken at a different photon energy. The data have 
	been normalized to the photon flux. The overall intensity 
	slope clearly shows the Cooper's minimum for the Sn 4d core level at 
	the lowest energies. The spectrum taken at 
	$h\nu \sim 245$~eV is shown in the inset panel together with the fit 
	to two doublets, after background correction.}
	\label{fig1}
\end{figure}

In Fig.~\ref{fig1}, we show the Sn 4d core level photoemission 
spectra taken at room temperature in normal emission geometry for photon energies between 
120 and 300~eV. In these experimental conditions, the photon energy dependence of the 
photoemission spectra (after normalization to the photoemission cross-section)
mainly reflects the vertical distance between the Sn adatom and 
the Ge atom directly underneath, thus enhancing the sensitivity to the vertical 
distortions of the adatom layer.\cite{floreano} 
In fact, the Sn vertical ripple is the main structural parameter 
involved in the phase transition, since it is expected to disappear in the 
high temperature phase if a displacive transition takes place.
A detailed description of the Sn 4d core level analysis in terms of two 
spin-orbit split doublets is given 
in Ref.~\cite{petaccia}, where high resolution spectra have been 
studied to determine the fitting parameters used for the successive 
analysis of the PED data taken in the LT phase.
Also in the $(\sqrt 3 \times \sqrt3)$R30$^{\circ}$
phase at 300~K, the Sn 4d core levels have a shape that cannot be fitted with a 
single spin-orbit split doublet.\cite{uhrberg,avila,lelay,uhrberg2} 
At least one additional
doublet is needed to improve appreciably the quality of the fit (see 
the inset panel of Fig.~\ref{fig1}).

\begin{figure}[tbp]
	\centering
	\includegraphics{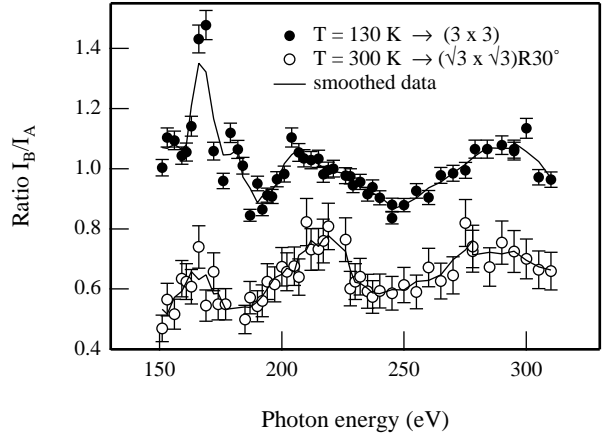}
	\caption{Intensity ratio between the minority $B$ and the majority $A$
components of the Sn 4d$_{5/2}$ core levels at 130~K (filled circles) and 300~K 
(open circles). The 
 data at 130~K have been vertically shifted by 0.4 units for the sake of clarity. 
 The smoothed data 
 (full lines) are also shown as a guide to the eye.}
	\label{fig2}
\end{figure}

The photon energy dependence of the intensity ratio $I_{B}/I_{A}$ between 
the minority $B$
and the majority $A$ components of the Sn 4d$_{5/2}$ core levels at 130~K and 300~K is 
shown in Fig.~\ref{fig2}. 
This presentation of the PED data allows a direct evaluation of 
any change of the Sn vertical ripple, in particular, the 
disappearance of the ripple would yield a constant intensity ratio 
without any energy modulation. In addition, the comparison between 
the intensity ratio below and above the transition is  
 not affected by the sistematic errors of extracting the 
anisotropy $\chi$-function. These errors can be large around 130~eV, 
because of the pronounced Cooper minimum. The error bars in 
Fig.~\ref{fig2} represent the maximum error in the intensity ratio, as 
obtained by the fitting procedure of ref.~\cite{petaccia}, and it 
mainly arises from the uncertainty on the energy shift between the 
$A$ and $B$ components and their linewidth. 
The intensity ratio $I_{B}/I_{A}$ below and above the phase transition is very 
similar, confirming that, also in the 
$(\sqrt 3 \times \sqrt3)$R30$^{\circ}$ phase at RT, there are two 
inequivalent types of Sn adatoms with different bonding geometry (different 
vertical distances between Sn and the underneath Ge atom). 
Thus, the rippled structure is maintained up to RT. 
This result differs from that obtained 
by SXRD experiments\cite{zhang,bunk} (where the surface corrugation 
was claimed to change from the $(3 \times 3)$ to the 
$(\sqrt 3 \times \sqrt3)$R30$^{\circ}$ phase), 
and indicates that the phase transition  at T$_{c} \sim$~220~K 
has an order-disorder character. 

\begin{figure}[tbp]
	\centering
	\includegraphics{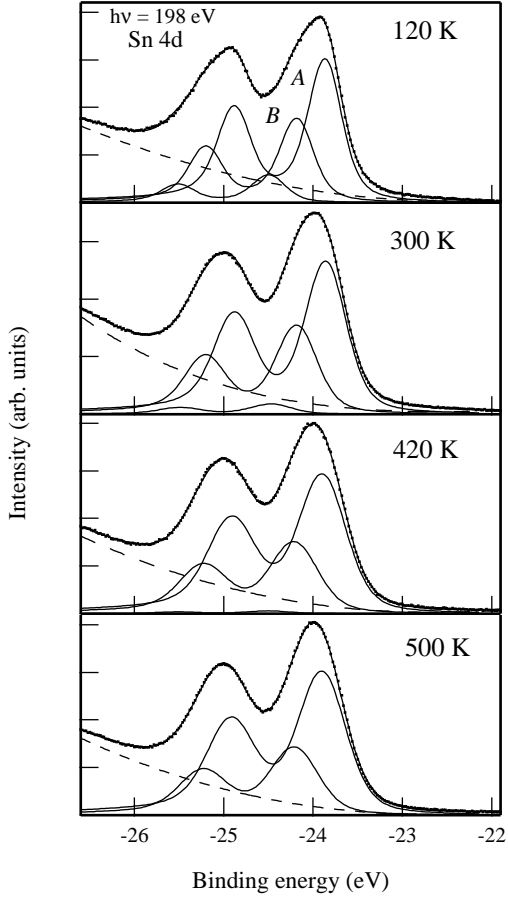}
	\caption{Fit of the Sn 4d core levels with three DS spin-orbit 
	split components in 
the $(3 \times 3)$-phase at 120 K, and in the 
$(\sqrt 3 \times \sqrt3)$R30$^{\circ}$-phase 
at 300, 420 and 500~K. The fits yield a Lorentzian width of 0.2~eV 
with a branching ratio of $0.63 \pm 0.01$ and an intensity ratio of 
0.52-0.6 between the $B$ and $A$ components.
}
	\label{fig3}
\end{figure}

Since the two spectral components are 
unambiguously assigned to two adsorption sites with different Sn heights, we can 
detect any structural modification of the 
$(\sqrt 3 \times \sqrt3)$R30$^{\circ}$ phase by simply measuring the Sn 4d 
spectrum as a function of the temperature. In particular, an eventual 
mergence of 
the two spectral components in a single one would be the fingerprint of a 
displacive transition.
We have reported in Fig.~\ref{fig3} the lineshapes of the Sn 4d core level spectra of 
1/3~ML of Sn/Ge(111) measured from 120~K to 500~K with a photon energy 
of 198~eV and an emission angle of 55$^{\circ}$~ from the surface normal. These spectra 
have been fitted to Doniach Sunjic (DS) doublets convoluted with a Gaussian. 
The Gaussian takes into account both the instrumental broadening (120~meV) 
and the temperature dependent phonon broadening, that is much larger. The free 
parameters were the energy position of the doublets, their intensities, 
the Gaussian 
full width at half maximum (FWHM) [assumed to be the same for all the 
doublets], 
the asymmetry of the DS line, and the branching ratio. The spin-
orbit splitting was set to 1.02~eV (according to 
Refs.~\cite{lelay,gothelid} and a polynomial function was used for the 
background caused by the low energy tail of the Ge 3d core levels at about 
29~eV of binding energy. 

\begin{figure}[tbp]
	\centering
	\includegraphics{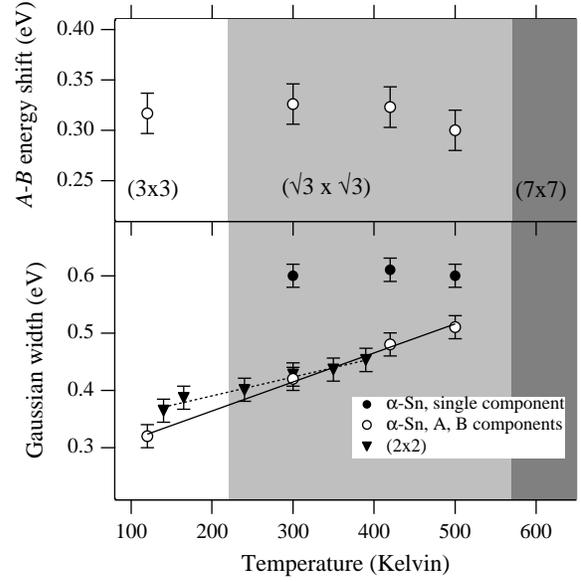}
	\caption{Upper panel: temperature dependence of the energy shift 
	between the $A$ and $B$ components as obtained by the fit to the 
	Sn 4d spectra in Fig.~3. 
Lower panel: temperature dependence of the Gaussian 
FWHM as obtained from Fig.~3 (open circles). 
The filled circles show the 
Gaussian FWHM obtained by fitting the spectra of the 
$\alpha$-phase to  a single spin-orbit split component (DS type). 
The Gaussian FWHM of the Sn 4d spectra, taken in the $(2 \times 2)$ 
phase at 0.2~ML, is also reported (filled triangles). Full and dotted 
lines are linear fits to the temperature behaviour of the FWHMs.}
	\label{fig4}
\end{figure}

Besides the $A$ and $B$ components, a third weak doublet was required to 
fit properly the $(3 \times 3)$ phase spectra at 120~K. This component 
is attributed to the Sn atoms surrounding Ge substitutional impurities 
in the $(3 \times 3)$ overlayer, according to the measurements of 
Uhrberg et al.\cite{uhrberg2}. We observed that the third component 
intensity decreases by increasing the temperature up to disappearance 
at 420~K. This is possibly due to limits of the fitting procedure in 
detecting such a faint feature when the main components become 
increasingly broader.\cite{uhrberg2}
 The energy shift between 
the $A$ and $B$ 
components does not change between 120~K and 500~K (see the upper panel of 
Fig.~\ref{fig4}), while the corresponding 
gaussian width gradually broadens from $\sim 0.3$~eV up to  $\sim 0.5$~eV.
The Sn 4d core level lineshape at 500~K is more symmetric and can be fitted to a 
single DS doublet, thus yielding a Gaussian FWHM of $\sim 0.6$~eV, a value much 
larger than that of the components used to fit the low temperature 
spectra. 

In order to 
discriminate between the model with a single component at 500~K and that with 
$A$ and $B$ components, we have followed the temperature dependence of the Gaussian 
FWHM of these components in the $(3 \times 3)$, in the $(\sqrt 3 \times 
\sqrt3)$R30$^{\circ}$ and in the $(2 \times 2)$ 
phases of Sn on Ge(111). 
 The corresponding temperature dependences are reported in Fig.~\ref{fig4}.
The values obtained for the $A$ and $B$ components of the $\alpha$-phase 
and for the $(2 \times 2)$ phase show a linear increase 
with the temperature, with a slope of 
$(0.5 \pm 0.02) \cdot 10^{-3}$~eV~K$^{-1}$ and 
$(0.32 \pm 0.03) \cdot 10^{-3}$~eV~K$^{-1}$, respectively. 
This result is consistent with the linear increase 
($\sim 0.2 \cdot 10^{-3}$~eV~K$^{-1}$) observed for the surface 
components of the 3d core 
levels of the clean Ge(111) surface in a wider 
temperature range.\cite{goldoni2}
The fit of the $\alpha$-phase core levels with a single component yields a 
gaussian width which does not change with the temperature, while the 
quality of the fit declines as the temperature is 
lowered (being definitely insufficient already at 300~K). 
If the 500~K spectra contained a single component, 
there would be a jump of 
$\sim 0.2$~eV of the FWHM between 300~K and 500~K, with about a threefold increase 
in the temperature derivative of the core level width. This 
implies that the thermal broadening of the Sn 4d core levels 
would be strongly non linear, in 
contrast to the data of the Sn $(2 \times 2)$ phase, the clean 
Ge(111) surface and other semiconductors 
surfaces.\cite{gothelid,paggel} 
As a consequence, the temperature dependence of the FWHMs reported in 
Fig.~\ref{fig4} 
strongly supports a deconvolution of the Sn 4d core level spectra 
with at least two DS doublets up to 500~K.

In conclusion, we have measured the temperature and energy dependence of 
the Sn 4d core levels of Sn/Ge(111). By means of energy dependent PED, we 
found that the vertical distortion, obtained for the two kinds of Sn adatoms in the 
LT $(3 \times 3)$ phase, is also present in the 
$(\sqrt 3 \times \sqrt3)$R30$^{\circ}$ phase at RT, i.e. the transition 
at $\sim$~220~K has an order-disorder character. In addition, we have shown that the 
two components of the photoemission spectra, reflecting the inequivalence of the 
Sn adsorption sites, are  present up to 500~K. This result leads us to discard the 
occurrence of a displacive character of the transition even at
temperatures much higher than T$_{c} = 220$~K.

This work was partly supported  
by MURST cofin99 (Prot. 9902332155 and Prot. 9902112831), 
by Regione Friuli-Venezia Giulia 98 and 
by INFM PAIS-F99.


\newpage
\onecolumn

\newpage
\twocolumn

\end{document}